\documentclass[11pt]{article}

\usepackage[utf8]{inputenc}
\usepackage{amsmath,amssymb,amsfonts,amsthm}
\usepackage{braket}
\usepackage{mathrsfs}
\usepackage{graphicx} 
\usepackage{bbold}
\usepackage{stmaryrd}
\usepackage{cite}
\usepackage{verbatim}
\usepackage{float}
\usepackage{color}
\usepackage[a4paper, total={6in, 9.5in}]{geometry}
\usepackage{bbm}
\usepackage{appendix}
\usepackage[colorlinks=true,allcolors=blue]{hyperref}
\usepackage{authblk}

\newcommand{\diag}{\mathrm{diag}}
\newcommand{\vb}[1]{{\boldsymbol{#1}}}
\newcommand{\Tr}[1]{\mathrm{Tr}#1}
\newcommand{\pderivative}[2]{\frac{\partial #1}{\partial #2}}

\begin{document}

\vspace*{2cm}
\begin{center}

{\LARGE\bf \vspace{0.2cm}
{Symmetry Restoration and Uniformly Accelerated Observers in Minkowski Spacetime}}
\vspace{1.5cm}

\textrm{\large Domenico Giuseppe Salluce,$^{a,e,}$\footnote{\texttt{salluce.domenicogiuseppe@spes.uniud.it}}
Marco Pasini,$^{a,c,e,}$\footnote{\texttt{marco.pasini@uniud.it}}
Antonino Flachi,$^{b,c,}$\footnote{\texttt{flachi@phys-h.keio.ac.jp}}\\
Antonio Pittelli,$^{d,f,}$\footnote{\texttt{antonio.pittelli@unito.it}}
Stefano Ansoldi$^{a,e,g,}$\footnote{\texttt{ansoldi@fulbrightmail.org}}}

\vspace{2em}

\vspace{1em}
\begingroup\itshape
${}^a$ Department of Mathematics, Computer Science, and Physics, University of Udine,\\ Via delle Scienze 206, I-33100 Udine, Italy
\\\vspace{0.5em} ${}^b$ Department of Physics, Keio University, 4-1-1 Hiyoshi, Yokohama,\\ Kanagawa 223-8521, Japan
\\\vspace{0.5em} ${}^c$  Research and Education Center for Natural Sciences, Keio University,\\ 4-1-1 Hiyoshi, Yokohama, Kanagawa 223-8521, Japan
\\\vspace{0.5em} ${}^d$ Dipartimento di Matematica, Universit\`a di Torino, Via Carlo Alberto 10,\\ I-10123 Torino, Italy
\\\vspace{0.5em} ${}^e$ Istituto Nazionale di Fisica Nucleare (INFN), Sezione di Trieste, Via Valerio 2,\\ I-34127 Trieste, Italy
\\\vspace{0.5em} ${}^f$ Istituto Nazionale di Fisica Nucleare (INFN), Sezione di Torino, Via Pietro Giuria 1,\\ I-10125 Torino, Italy
\\\vspace{0.5em} ${}^g$ Institute for Fundamental Physics of the Universe (IFPU), Via Beirut 2,\\ I-34151 Trieste, Italy
\par\endgroup

\end{center}

\vspace{1em}

\begin{abstract}\noindent

 We reassess the problem of symmetry restoration induced by observers' acceleration within the context of interacting quantum field theories in Minkowski spacetime. We argue that the imposition of a frame-independent renormalization condition negates any observed symmetry restoration by a Rindler observer. Technically, we compute the one-loop effective potential of a $\lambda\varphi^4$ theory for an accelerated observer, employing a distinct methodology from prior investigations. Emphasizing the intricacies of the model's renormalization, the analysis offers novel insights into the interplay between acceleration and spontaneous symmetry breaking in quantum field theory.

\end{abstract}

\newpage

\overfullrule=0pt
\parskip=2pt
\parindent=12pt
\headheight=0.0in \headsep=0.0in \topmargin=0.0in \oddsidemargin=0in

\vspace{-3cm}
\vspace{-1cm}

\tableofcontents

\setcounter{footnote}{0}

\section{Introduction}
  
The inherent non-uniqueness of canonical quantization for fields on Riemannian spacetime leads to an ambiguity in defining a fundamental vacuum state. This ambiguity arises due to the intricate relationship between the choice of spacetime coordinates and their association with the Hamiltonian in relativistic quantum field theory. Even in  Minkowski spacetime, the traditional vacuum state ceases to be the ground state when observed in a non-inertial coordinate system \cite{Fulling:1972md}. As a consequence, in adopting uniformly accelerated coordinates, the ordinary Minkowski vacuum is no longer void but encompasses radiation. Ideally, such a phenomenon might be probed by a  detector experiencing acceleration within the vacuum, registering a non-empty state due to a  radiation exhibiting a thermal spectrum. Quantitatively, an accelerated idealized thermometer records a non-vanishing temperature $T_U$ proportional to the acceleration $a$ (measured from an inertial rest frame), expressed by the equation
\begin{equation}
    T_U = \frac{a \hbar}{2 \pi k c},
    \label{TU}
\end{equation}
where, as usual,  $\hbar$ denotes the reduced Planck constant, $k$   Boltzmann's constant  and $c$   the speed of light\footnote{For the sake of notational convenience, subsequent calculations consider $\hbar = c = k = 1$, simplifying the expression to $T_U = a / (2 \pi)$.}. This observation fundamentally highlights that the temperature of a system undergoing acceleration in vacuum experiences a substantial increase due to interactions with quantum fluctuations. This phenomenon encapsulates the core of the Unruh effect, with $T_U$ defining the Unruh temperature \cite{Unruh:1976db}.

The intriguing similarity between acceleration and temperature has opened avenues for investigating the interplay among gravity, thermodynamics and quantum field theory. This connection has spurred the notion that elementary particles could serve as effective radiation detectors. An early proposal ventured to relate the polarization effects of electrons circulating in a magnetic field, akin to uniform circular accelerated motion, with the temperature effect outlined in the Unruh effect \cite{Bell:1986ir,Bell:1984sr,Bell:1982qr}. This proposal sought to interpret the incomplete polarization of electrons within a storage ring through the lens of the Unruh effect.

The concept of utilizing elementary particles as thermometers takes a  turn when considering spontaneous symmetry breaking within the mechanism of \emph{heating by acceleration in vacuum}. The prospect arises from the realization that thermal effects can potentially trigger the restoration of a spontaneously broken symmetry \cite{Dolan:1973qd}. By naively extending this idea, the analogy between acceleration and temperature might suggest the possibility that acceleration, a  \emph{proxy} of temperature in the Unruh effect, could induce a transition between a broken and a restored symmetry phase for a specific critical acceleration and fixed interaction strength. 

Building upon this intuitive concept, investigating the potential induction of a transition within an interacting system through acceleration—akin to the process of melting a condensate—emerges as a pivotal inquiry in elucidating the intricate relationship between quantum field theory and general relativity. In particular, this hypothesis implies that a condensate might operate as an effective thermometer, offering a means to measure the critical acceleration of the associated phase transition. However, this intriguing proposition has been met with varied conclusions over the years, following the seminal work by Ref.~\cite{Unruh:1983ac}. Subsequent studies have revisited this question with differing outcomes~\cite{Casado-Turrion:2019gbg, Dobado:2017xxb, Benic:2015qha,Takeuchi:2015nga,Takeuchi:2014rba,Castorina:2012yg,Lenz:2010vn,Ebert:2008zza,Ebert:2006bh,Ohsaku:2004rv,Hill:1986ec,Hill:1985wi,Paredes:2008cr,Peeters:2007ti}.

The purpose of this work is to address this problem by means of a covariant renormalization scheme. The aim is to scrutinize the relationship between acceleration, thermodynamics, and quantum field theory, probing the potential of accelerated systems as thermodynamic analogs while grappling with the implications and varying conclusions drawn from extensive investigations in this domain. 
 In particular, we consider a scalar theory where spontaneous symmetry breaking is present according to an inertial-frame description; we rephrase the description according to the perspective of a uniformly accelerated observer, wondering if, for sufficiently high proper acceleration, the thermal bath predicted by Unruh is able to restore the broken symmetry of the system. We pass through the comparison of two different approaches, already present in the scientific literature, which seemingly lead to different answers ~\cite{Unruh:1983ac, Castorina:2012yg}. By retracing one of the two procedures with some variations, we show that it can be reconciled with the other.  

\paragraph{Outline.}

The plan of the paper is as follows. In Section~\ref{sec2} we remind some basic facts about the quantization of a free scalar field in a uniformly accelerated frame. These results set up our notation and are a useful starting point for the discussion of the interacting case to be addressed in Section~\ref{sec3}. In particular, in  Section~\ref{sec31}  we review the original result concerning the invariance of the vacuum ($n$-point) Green's functions presented in Ref.~\cite{Unruh:1983ac}, from which it follows straightforwardly that symmetry cannot be restored as an effect of acceleration, at least for the set-up that we consider in this paper. Furthermore, in  Section~\ref{sec32} we inspect the same problem using the effective action formalism and using a different implementation than those discussed before. Renormalization is one of the crucial points of our approach and it is discussed in detail in Section~\ref{sec4}, where we recover the conclusion of Ref.~\cite{Unruh:1983ac} using a covariant renormalization scheme. Our conclusions are that, at least for the system that we consider in this work, acceleration does not trigger a restoration of symmetry. This result is clarified in Sec.\ref{sec5} from the point of view of a covariant, point-splitting approach to regularization. We conclude with some comments and remarks in section~\ref{sec6}.

\section{Rindler quantization in a free-field scalar theory\label{sec2}}

In this work, we are interested in an interacting relativistic scalar field theory described by the action
\begin{equation}
    S=\int d^4x\biggl(\frac{1}{2}\partial^\mu\phi\,\partial_\mu\phi - \frac{1}{2}m^2\phi^2 - \frac{\lambda}{4!}\phi^4\biggr).
    \label{ssblag}
\end{equation}
The underlying background geometry is assumed to be Minkowski spacetime with signature $(+,\,-,\,-,\,-)$ and $x^\mu=(t,\,x,\,\vb{x}_\perp)$, where $\vb{x}_\perp=(y,\,z)$, represents the conventional Cartesian coordinates.

We begin assuming the system described by the action (\ref{ssblag}) to be at rest in an inertial reference frame with $m^2<0$, meaning that in the absence of acceleration, the theory is in a broken symmetry phase at zero temperature. 
In particular, the action \eqref{ssblag} is invariant under the $\mathbb{Z}_2$-transformation $\phi\rightarrow - \phi$, but the vacuum expectation value $\langle\phi(x)\rangle$ is not. Therefore, in the quantum theory there must be two degenerate vacuum states, $|0_{M+}\rangle$ and $|0_{M-}\rangle$. To the tree-level one has that
 \begin{equation}
     \bra{0_{M\pm}}\phi(x)\ket{0_M\pm}\simeq \varphi^{\pm}_{cl},
 \end{equation} 
 where
\begin{equation}
    \varphi^{\pm}_{cl}=\pm\biggl(-\frac{6\,m^2}{\lambda}\biggr)^{1/2}
\end{equation}
 are the minima of the potential
 $V(\varphi)=\frac{1}{2}m^2\varphi^2 + \frac{\lambda}{4!}\varphi^4$. 
The two vacua are such that $\ket{0_{M+}}\leftrightarrow\ket{0_{M-}}$ under $\phi\rightarrow-\phi$, and are equivalent in principle, but only one of the two is chosen when the theory is perturbed in a $\mathbb{Z}_2$-breaking way. Details can be found e.g. in ~\cite{Weinberg:1996kr}. 

The broken symmetry can in principle be restored at high temperature, e.g. by putting the system in a thermal bath at a temperature $T$ larger than a critical one $T_{cr}$; in this regime, one has $\Tr\,[\phi(x)\,e^{-H/T}]/\Tr\,[e^{-H/T}]=0$, being $H$ the Hamiltonian of the system \cite{Dolan:1973qd}.

The question that we wish to address is whether an accelerated observer (who sees the Minkowski vacuum as a thermal state) can see a transition into a phase of restored symmetry, provided that his/her proper acceleration exceeds a critical value. 

To describe the perspective of a uniformly accelerated observer with proper acceleration $a$, we perform a coordinate transformation to Rindler coordinates $\bar{x}^\mu=(\tau,\,\rho,\,\vb{x}_\perp)$ 
\begin{equation}
   \begin{cases}
  t=\phantom{-}\rho\,\sinh{a\tau}, \quad x=\phantom{-} \rho\, \cosh{a\tau}, \quad \rho>0 \quad (R) \\
   t= -\rho\,\sinh{a\tau}, \quad x= -\rho\, \cosh{a\tau}, \quad \rho>0 \quad (L) \\
\end{cases}
\label{rindcoordtrasf}
\end{equation}
where the right ($R$) and left ($L$) wedges are defined in Minkowski spacetime \cite{Sciama:1981hr,Takagi:1986kn,Crispino:2007eb}.
\begin{figure}
    \centering
    \includegraphics{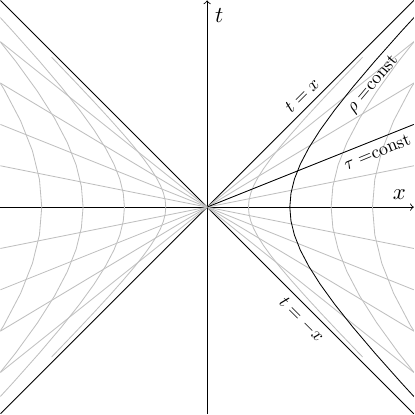}
    \caption{Rindler wedges in the plane $(t,\,x)$. Rindler coordinates are defined in the regions $R$ and $L$: time coordinates $\tau=\mathrm{const}$ are straight lines passing through the origin, and space coordinates $\rho=\mathrm{const}$ are branches of hyperbolae, corresponding to the worldlines of different Rindler observers.}
    \label{fig1}
\end{figure}
We shall refer to the \textit{union} $R\cup L$ as the \textit{Rindler space} (Figure \ref{fig1}).
The line element in these coordinates takes the following static form
\begin{equation}
    ds^2=\rho^2\,a^2\,d\tau^2-d\rho^2-dy^2-dz^2.
    \label{rindlermetric}
\end{equation}
Before moving on to the interacting case, we shall briefly discuss the quantization of the \textit{free theory} (action \eqref{ssblag} with $\lambda=0$) in order to establish some elementary results to be used later. The Klein-Gordon equation corresponding to Rindler coordinates is
\begin{equation}
    \biggl(\frac{1}{(\rho\,a)^2}\partial^2_\tau-\frac{1}{\rho}\partial_\rho\,\rho\,\partial_\rho-\partial^2_y-\partial^2_z+m^2 \biggr)\phi=0
    \label{kgrindler}.
\end{equation}
Solutions to Eq.~\eqref{kgrindler} can be found in the right $R$ wedge in the form 
\begin{equation}
    f^R_{\omega\vb{k}_{\perp}}=\frac{1}{(2\pi)^2}\biggl[\frac{4}{a}\,\sinh(\frac{\omega\pi}{a})\biggr]^\frac{1}{2}\,K_{i\frac{\omega}{a}}(\mu_{\vb{k}_{\perp}}\rho)e^{-i\omega\tau}\,e^{i\vb{k}_\perp\cdot\vb{x}_\perp}
    \label{normrindmodes}
\end{equation}
where $K_\nu$ is the modified Bessel function of the second kind, and $\mu_{\vb{k}_{\perp}}=\sqrt{\vb{k}_\perp^2+m^2}$. These modes \eqref{normrindmodes} are orthonormal with respect to the Klein-Gordon inner product, which, given two solutions $f_1$ and $f_2$ of~\eqref{kgrindler}, can be written as
\begin{align}
 (f_1,\,f_2)\equiv i\int_{\rho>0}\,\frac{d\rho}{|\rho\,a|}\, d^2x_\perp (f_1^*\partial_\tau f_2-f_2\partial_\tau f_1^*),  
  \label{kgscpgen}
\end{align}
in Rindler coordinates over $R$. 

Canonical quantization can be carried out in the standard way by expanding the field operator in terms of the set $\{f^R_{\omega\vb{k}_{\perp}}\}\cup\{f^{R*}_{\omega\vb{k}_{\perp}}\}$, which is complete over the wedge $R$:
\begin{equation}
    \phi=\int_{\omega>0} d\omega\,d^2k_{\perp}(a^R_{\omega\vb{k}_{\perp}}\,f^R_{\omega\vb{k}_{\perp}}+{a^{R\dag}_{\omega\vb{k}_{\perp}}}\,{f_{\omega\vb{k}_{\perp}}^{R*}}).
    \label{rindlerexpR}
\end{equation}
Imposing the following commutation relations between the (annihilation and creation) operators $a^R_{\omega\vb{k}_{\perp}}$ and $a^{R\dag}_{\omega'\vb{k}'_{\perp}}$,
\begin{equation}
\begin{aligned}
      \relax[a^R_{\omega\vb{k}_{\perp}},\,a^R_{\omega'\vb{k}'_{\perp}}]&=[a^{R\dag}_{\omega\vb{k}_{\perp}},\,a^{R\dag}_{\omega'\vb{k}'_{\perp}}]=0,\\ 
      [a^R_{\omega\vb{k}_{\perp}},\,a^{R\dag}_{\omega'\vb{k}'_{\perp}}]&=\delta(\omega-\omega')\,\delta^{(2)}(\vb{k}_{\perp}-\vb{k}'_{\perp}), 
      \label{commrelrind}
\end{aligned}
\end{equation}
allows one to define the (\textit{Fulling-Rindler}) vacuum state $\ket{0_R}$ in the $R$ wedge as follows:
\begin{equation}
    a^R_{\omega\vb{k}_{\perp}}\ket{0_R}=0 \quad \forall\,\omega,\,\vb{k}_{\perp}.
\end{equation}
The quantization procedure can be carried out similarly over the left wedge, $L$, by expanding the field in modes\footnote{Here modes $f^L_{\omega\vb{k}_{\perp}}$ have the same form as in~\eqref{normrindmodes}.} as
\begin{equation}
\phi=\int_{\omega>0} d\omega\,d^2k_{\perp}(a^L_{\omega\vb{k}_{\perp}}\,{f^{L*}_{\omega\vb{k}_{\perp}}}+{a^{L\dag}_{\omega\vb{k}_{\perp}}}\,{f^L_{\omega\vb{k}_{\perp}}}),
\label{rindlerexpL}
\end{equation}
and by imposing commutation relations between the $a^L_{\omega\vb{k}_{\perp}}$ and $a^{L\dag}_{\omega'\vb{k}'_{\perp}}$ operators analogous to the~\eqref{commrelrind}. This allows one to define another vacuum state $\ket{0_L}$ in the wedge $L$, which satisfies
\begin{equation}
     a^L_{\omega\vb{k}_{\perp}}\ket{0_L}=0 \quad \forall\,\omega,\,\vb{k}_{\perp}.
\end{equation}
The union of the two sets of $L$-modes and $R$-modes is complete over the whole Rindler space $R\cup L$ (here the $L/R$-modes are assumed to vanish in the region $R/L$, respectively), with the corresponding expansion taking the form:
\begin{equation}
    \phi=\int_{\omega>0} d\omega\,d^2k_{\perp}( a^R_{\omega\vb{k}_{\perp}}\,f^R_{\omega\vb{k}_{\perp}}+{a^{R\dag}_{\omega\vb{k}_{\perp}}}\,{f_{\omega\vb{k}_{\perp}}^{R*}} 
    +a^L_{\omega\vb{k}_{\perp}}\,{f^{L*}_{\omega\vb{k}_{\perp}}}+{a^{L\dag}_{\omega\vb{k}_{\perp}}}\,{f_{\omega\vb{k}_{\perp}}^{L}}).
\label{rindlerexp}
\end{equation}

We explicitly remark that \emph{Rindler modes do not share the same analyticity properties as the standard Minkowski modes} used in inertial frames: this implies that the two quantization procedures lead to inequivalent particle structures (or unitarily inequivalent constructions)~\cite{Fulling:1972md,Unruh:1976db}. However, it is possible to construct two independent linear combinations of Rindler modes with the appropriate analyticity properties~\cite{Unruh:1976db}; these take the following normalized expressions
\begin{align}
     f^{A}_{\omega\vb{k}_{\perp}}=\frac{1}{|2\sinh(\frac{\omega\pi}{a})|^{1/2}}(e^{\frac{\pi\omega}{2a}}f^R_{\omega\vb{k}_{\perp}}+e^{-\frac{\pi\omega}{2a}}f^L_{\omega\vb{k}_{\perp}}),
    \label{analytic} \\
    f^B_{\omega\vb{k}_{\perp}}=\frac{1}{|2\sinh(\frac{\omega\pi}{a})|^{1/2}}(e^{-\frac{\pi\omega}{2a}}{f^{R*}_{\omega\vb{k}_{\perp}}}+e^{\frac{\pi\omega}{2a}}{f^{L*}_{\omega\vb{k}_{\perp}}}) 
     \label{analyticbis}
\end{align}
with $\omega>0$. Using these $A$ and $B$ combinations as basis, one can express the decomposition \eqref{rindlerexp} as:
\begin{equation}
     \phi=\int_{\omega>0} d\omega\,d^2k_{\perp}( a^A_{\omega\vb{k}_{\perp}}\,f^A_{\omega\vb{k}_{\perp}}+{a^{A\dag}_{\omega\vb{k}_{\perp}}}\,{f_{\omega\vb{k}_{\perp}}^{A*}}
    +a^B_{\omega\vb{k}_{\perp}}\,{f^{B}_{\omega\vb{k}_{\perp}}}+{a^{B\dag}_{\omega\vb{k}_{\perp}}}\,{f_{\omega\vb{k}_{\perp}}^{B*}}).
    \label{analyticexpansion}
\end{equation}
This construction is unitarily equivalent (i.e., it leads to an equivalent quantization) to the standard inertial one, i.e. its annihilation operators define a vacuum state $\ket{0_M}$ corresponding to the one that we obtain in Minkowski spacetime. Moreover, the $A$-operators commute with the $B$-operators defined in~(\ref{analyticexpansion}), and the two sets separately satisfy commutation relations that are analogous to~\eqref{commrelrind}.

\section{Interacting quantum field theories\label{sec3}}

We can now examine what happens in the presence of interactions. First, we briefly remind the reader of the original result of Ref.~\cite{Unruh:1983ac}, whose conclusions will be later recovered using a different approach.

\subsection{Green function approach\label{sec31}}

The generalization of the Unruh effect for an interacting quantum field theory was first examined by Sewell in Ref. ~\cite{Sewell:1982zz} with an axiomatic scheme, for a class of manifolds including Schwarzschild spacetime as well as Minkowski spacetime, and later by Unruh and Weiss in \cite{Unruh:1983ac} in the path integral approach. The latter work considers both scalars and fermions with a polynomial interaction; here, without restrictions, we only consider the case of scalars.
Let us first consider the following \textit{Rindler} Hamiltonian restricted to the right wedge
\begin{equation}
    H^{\mathcal{R}}=\int_{\rho>0} d\rho\, d^2x_\perp\,(\rho\,a)\biggl(\frac{(\pi^\mathcal{R})^2}{2}+\frac{(\partial_\rho\phi)^2}{2}+\frac{(\nabla_\perp\phi)^2}{2}+V(\phi)\biggr)
    \label{rindler_hamiltonian}
\end{equation}
where $\pi^{\mathcal{R}}$ is the canonically conjugate momentum to $\phi$. The Hamiltonian \eqref{rindler_hamiltonian} is the generator of $\tau$-translations in $R$ (it could be analogously defined for the left wedge $L$, up to a sign change due to the fact that the vector field $\partial_\tau$ is past-directed in $L$). This is not the usual Minkowski Hamiltonian (the one studied e.g. in chapter 2 of \cite{Peskin:1995ev}), but it corresponds, instead, to $a$ times the generator of Lorentz-boosts, restricted to the right Rindler wedge. The main result obtained in Ref.~\cite{Unruh:1983ac} is the proof of the following equality
\begin{equation}
    \bra{0_M}(\phi(x_1)...\phi(x_n))_t\ket{0_M}=\frac{\Tr\small[e^{-\beta_U\, H^\mathcal{R}} (\phi(\bar{x}_1)...\phi(\bar{x}_n))_\tau]}{\Tr\small[e^{-\beta_U\, H^\mathcal{R}}]},
    \label{unruhweiss}
\end{equation}
where $\beta_U = T_U^{-1}$. In \eqref{unruhweiss} the points $x_i$ have to be within the same Rindler wedge (e.g., $R$), and so does the trace $\Tr\,$. This technical subtlety is also emphasized in \cite{Sewell:1982zz}, and it is an important detail in the parallel between relations like \eqref{unruhweiss} and the idea of thermofield dynamics \cite{Takahasi:1974zn}, as first noted in \cite{Israel:1976ur}.
Moreover, the time ordering in \eqref{unruhweiss} can be taken, \emph{equivalently}, with respect to $t$ or $\tau$. What Eq.~(\ref{unruhweiss}) implies is that all vacuum Green's functions between spacetime points in the same Rindler wedge in Minkowski coordinates are the same as the real-time Green's functions of the Rindler observer in thermal equilibrium at a temperature $T_U$ \cite{Unruh:1983ac}: physically, the accelerated observer sees a thermal spectrum at the Unruh temperature in the Minkowski vacuum state. From this result, Ref.~\cite{Unruh:1983ac} argues that, given a theory with multiple vacua and where a phase transition is expected to occur as a function of temperature, the same does not occur as a function of acceleration (Ref.~\cite{Unruh:1983ac} considers as an example a scalar theory of the form (\ref{ssblag})). Concretely, this is done by assuming that the $\lambda \phi^4$ theory \eqref{ssblag}, which has two degenerate vacua, is perturbed by a symmetry-breaking term $J$, in order for the resulting ground state to be unique (i.e., it is a unique vacuum $\ket{0_M}$). Hence, an inertial observer measures
\begin{equation}
    \bra{0_M}\phi(x)\ket{0_M}\neq0.
    \label{inertialVev}
\end{equation}
The above is a statement that for an inertial observer, symmetry is spontaneously broken. The question is whether, for a Rindler (accelerated) observer with proper acceleration $a$, the symmetry can be restored for some sufficiently large value of the acceleration. This is the case if
\begin{equation}
    \Tr[e^{-\beta_U\,H^{\mathcal{R}}}\phi(\bar{x})]/\Tr[e^{-\beta_U H^{\mathcal{R}}}]=0.
\end{equation}
However, from~\eqref{unruhweiss} and from~(\ref{inertialVev}) it follows that
\begin{equation}
    \Tr[e^{-\beta_U\,H^{\mathcal{R}}}\phi(\bar{x})]/\Tr[e^{-\beta_U H^{\mathcal{R}}}]=\bra{0_M}\phi(x)\ket{0_M}\neq0.
\end{equation}
The above relation has to be understood in the limit of a vanishing perturbation $J$. Hence, Ref.~\cite{Unruh:1983ac} concludes that there is no threshold value of $a$ beyond which the Rindler observer detects a symmetry restoration. The same conclusion is reached, for example, in Ref.~\cite{Hill:1985wi,Hill:1986ec}.

\subsection{Effective action approach\label{sec32}}

An alternative way to address the same problem is by computing the effective potential, as we describe below. This approach has been taken in a number of works, see for example Refs.~\cite{Castorina:2012yg, Ebert:2008zza,Ebert:2006bh,Ohsaku:2004rv}, where a different conclusion is reached.

We shall take the point of view of an accelerated observer located in the $R$ wedge, who measures a (Unruh) temperature locally proportional to his/her own proper acceleration. The action of our system as seen by the accelerated observer (in Rindler coordinates) is
\begin{equation}
    S^{\mathcal{R}}=\int d^4\bar{x}\sqrt{-g}\biggl(g^{\mu\nu}\frac{1}{2}\,\partial_\mu\varphi\,\partial_\nu\varphi-\frac{1}{2}m^2\varphi^2-\frac{\lambda}{4!}\varphi^4\biggr)
    \label{ssbacc}
\end{equation}
with
\begin{equation}
\begin{aligned}
     &g_{\mu\nu}=\diag(\rho^2a^2,\,-1,\,-1,\,-1), \\
     &g=\det g_{\mu\nu}=-\rho^2a^2.
\end{aligned}
\end{equation}
The difference between the above action $S^{\mathcal{R}}$ and expression~(\ref{ssblag}) is the explicit presence in~(\ref{ssbacc}) of the metric tensor. Starting from~(\ref{ssbacc}), we can (i) evaluate the one-loop effective action $\Gamma^{\mathcal{R}}$ at finite (Unruh) temperature by expanding the action around a constant background field $\varphi_0$, and (ii) compute the quadratic path integral with periodic boundary conditions, $\varphi(\tau-i\beta_U)= \varphi(\tau)$, which are consistently imposed on the field in a scalar thermal state at temperature $T_U$. This leads to the well-known general expression 
\begin{equation}
    \Gamma^{\mathcal{R}}(\varphi_0)=S^{\mathcal{R}}(\varphi_0)+\frac{i}{2}\Tr\ln[\partial_{\mu}\sqrt{-g}g^{\mu\nu}\partial_{\nu}+\sqrt{-g}M^2] ,
    \label{rindlereffact}
\end{equation}
where $M^2=m^2+\lambda\varphi_0^2/2$. The trace of the differential operator can be evaluated by spanning the related Hilbert space in terms of a basis of position eigenstates $\ket{x}$ (see e.g. ~\cite{Christensen:1976vb} for reference), satisfying the Dirac orthonormality property $\braket{x|x'}=\delta(x-x')$, and the completeness relation $\mathbb{1} =\int\,d^4x \ket{x}\bra{x}$. Thus, it is possible to rewrite the 1-loop contribution to the effective action as follows
\begin{equation}
\begin{aligned}
\label{new_trln}
   & \Tr\ln[\partial_{\mu}(\sqrt{-g}g^{\mu\nu}\partial_{\nu})+\sqrt{-g}M^2]\\
  &  =\int d^4\bar{x}' \, \int_0^{M^2}\,dq \frac{d}{dq}\left(\bra{\bar{x}'}\ln\left[\partial_\mu\sqrt{-g}g^{\mu\nu}\partial_\nu+\sqrt{-g}q\right]\ket{\bar{x}'}\right)+\int d^4\bar{x}'\bra{\bar{x}'}\ln\left[\partial_\mu\sqrt{-g}g^{\mu\nu}\partial_\nu\right]\ket{\bar{x}'},
\end{aligned}
\end{equation}
or equivalently, using the fact that $G(\bar{x},\,\bar{x}',\,m)=\bra{\bar{x}}\left[\partial_\mu\sqrt{-g}g^{\mu\nu}\partial_\nu+\sqrt{-g}M^2\right]^{-1}\ket{\bar{x}'}$, equation (\ref{new_trln}) takes the form
\begin{equation}
\begin{aligned}
 &  \Tr\ln[\partial_{\mu}(\sqrt{-g}g^{\mu\nu}\partial_{\nu})+\sqrt{-g}M^2]\\
&=\int d^4\bar{x}' \sqrt{-g}\, \int_0^{M^2}\,dq\, \lim_{\bar{x}\rightarrow\bar{x}'}G(\bar{x},\,\bar{x}',\,m)+\int d^4\bar{x}'\bra{\bar{x}'}\ln\left[\partial_\mu\sqrt{-g}g^{\mu\nu}\partial_\nu\right]\ket{\bar{x}'},
    \label{tracerindler2}
\end{aligned}
\end{equation}
where
\begin{equation}
    G(\bar{x},\,\bar{x}',\,m)\equiv i\,\Tr\small[e^{-\beta_U\,H^{\mathcal{R}}}(\phi(\bar{x})\,\phi(\bar{x}'))_{\tau}]/\Tr\small[e^{-\beta_U H^{\mathcal{R}}}]|_{\text{free theory}}
\end{equation}
is the free-field thermal propagator in the $R$ wedge. In the RHS of \eqref{tracerindler2}, the integral of the logarithm is a purely kinetic term, therefore independent of $\varphi_0^2$, so it can be neglected since we are interested in the $\varphi_0^2$ differentiation of the effective potential. From the definition of the effective potential $V^{\mathcal{R}}(\varphi_0)$,
\begin{equation}
    \Gamma^{\mathcal{R}}(\varphi_0)=-\int\,d^4\bar{x}\sqrt{-g}\,V^{\mathcal{R}}(\varphi_0),
\end{equation}
and using \eqref{tracerindler2} in~\eqref{rindlereffact}, we obtain
\begin{equation}
    V^{\mathcal{R}}(\varphi_0)=\frac{1}{2}m^2\varphi_0^2+\frac{\lambda}{4!}\varphi_0^4-\frac{i}{2}\int_0^{M^2}\,dq^2\lim_{\bar{x}\rightarrow\bar{x}'}G(\bar{x},\,\bar{x}',\,q) .
    \label{effpotgeneralcf}
\end{equation}
Knowledge of the effective potential allows us to obtain the necessary conditions for symmetry restoration: the coefficient of the quadratic term in the effective action must be positive (with $\lambda \geq 0$), i.e.,
\begin{equation}
    0 \leq \pderivative{V^{\mathcal{R}}}{\varphi^2_0}\bigg|_{\varphi_0=0} .
    \label{neccondition}
\end{equation}
In our case, we obtain 
\begin{equation}
    \pderivative{V^{\mathcal{R}}}{\varphi_0^2}\bigg|_{\varphi_0=0}=\frac{m^2}{2}-i\frac{\lambda}{4}\,\lim_{\bar{x}\rightarrow\bar{x}'}\,G(\bar{x},\,\bar{x}',\,q)
    \label{restcond}.
\end{equation}
Up to this point, we have closely followed the procedure discussed in the existing literature, and the result agrees with existing calculations. However, in the following we depart from previous calculations and choose to carry out the computation of the Green's function using the canonical formalism:
we, then, take the expansion of the field operator in terms of the analytic modes~\eqref{analyticexpansion},
where, as we know, $a^A_{\omega\,\vb{k}_\perp}$ and $a^{B}_{\omega\,\vb{k}_\perp}$ annihilate $\ket{0_M}$. 
As long as the equality~\eqref{unruhweiss} holds, we should have 
\begin{equation}
     G(\bar{x},\,\bar{x}',\,m)=i\,\bra{0_M}(\phi(\bar{x})\,\phi(\bar{x}'))_{\tau}\ket{0_M}.
     \label{feynmanmink}
\end{equation}
(In the case $\tau>\tau'$, the Feynman propagator is reduced to $\bra{0_M}\phi(\bar{x})\,\phi(\bar{x}')\ket{0_M}$.)
Upon substitution of the mode expansion~\eqref{analyticexpansion}, the only nonzero contributions in the RHS of~\eqref{feynmanmink} come from the terms containing $ a^A_{\omega\,\vb{k}_\perp}a^{A\dag}_{\omega\,\vb{k}_\perp}$ or $a^B_{\omega\,\vb{k}_\perp}a^{B\dag}_{\omega\,\vb{k}_\perp}$. Using the commutation rules for these operators, we have
\begin{equation}
\bra{0_M}a^A_{\omega\,\vb{k}_\perp}a^{A\dag}_{\omega\,\vb{k}_\perp}\ket{0_M}=\bra{0_M}a^B_{\omega\,\vb{k}_\perp}a^{B\dag}_{\omega\,\vb{k}_\perp}\ket{0_M}=
\delta(\omega-\omega')\delta^{(2)}(\vb{k}_\perp-\vb{k}'_\perp).
\label{nonnull}
\end{equation}
Using the above relations, along with (\ref{analytic}),(\ref{analyticbis}), (\ref{analyticexpansion}), in the Feynman propagator~(\ref{feynmanmink}) yields
\begin{equation}
\begin{aligned}
     G(\bar{x},\,\bar{x}',\,m)=&i\,\int_0^\infty \frac{d\omega}{a\,\pi^2}\,\cosh(\frac{\pi\omega}{a}-i\omega|\tau-\tau'|)\times \\
    &\int\,\frac{d^2k_\perp}{(2\pi)^2}\,\,e^{i\,\vb{k}_{\perp}\cdot(\vb{x}_{\perp}-\vb{x}'_{\perp})}\,K_{i\frac{\omega}{a}}(\mu_{\vb{k_{\perp}}}\rho)K_{i\frac{\omega}{a}}(\mu_{\vb{k_{\perp}}}\rho').
\end{aligned}
   \label{finalpropmink}
\end{equation}
This result is consistent with \cite{Linet:1995mq}. Similarly, we compute the Feynman propagator evaluated on the Rindler vacuum,
\begin{equation}
    G_{0}(\bar{x},\,\bar{x}',\,m)\equiv i\,\bra{0_R}(\phi(\bar{x})\,\phi(\bar{x}'))_{\tau}\ket{0_R}.
     \label{feynmanrind}
\end{equation}
The procedure is analogous, but here we expand the field operators on the Rindler modes $f^{R}_{\omega\vb{k}_\perp}$ using \eqref{rindlerexpR}, which leads to
\begin{equation}
\begin{aligned}
     G_0(\bar{x},\,\bar{x}',\,m)=&i\,\int_0^\infty \frac{d\omega}{a\,\pi^2}\,\sinh(\frac{\pi\omega}{a})\,e^{-i\omega|\tau-\tau'|}\times \\
    &\int\,\frac{d^2k_\perp}{(2\pi)^2}\,\,e^{i\,\vb{k}_{\perp}\cdot(\vb{x}_{\perp}-\vb{x}'_{\perp})}\,K_{i\frac{\omega}{a}}(\mu_{\vb{k_{\perp}}}\rho)K_{i\frac{\omega}{a}}(\mu_{\vb{k_{\perp}}}\rho'),
\end{aligned}
   \label{finalproprind}
\end{equation}
and is consistent with, both,~\cite{Linet:1995mq} and~\cite{Candelas:1976jv}.

In order to obtain the effective potential, we need to compute the coincidence limit $\bar{x}\rightarrow\bar{x}'$ of the Green's function. In this limit the two propagators  $G_0(\bar{x},\,\bar{x}',\,m)$ and  $G(\bar{x},\,\bar{x}',\,m)$ both diverge, as it can be seen by performing the integration in $d^2k_\perp$. This can be explicitly calculated in the limit $m\rho\ll 1$, leading to
\begin{align}
    \lim_{\bar{x}\rightarrow\bar{x}'}G(\bar{x},\,\bar{x}',\,m)=&\frac{i}{(2\pi)^2}\frac{1}{(a\rho)^2}\,\int_0^\infty \,d\omega\,\omega\,\coth(\frac{\pi\omega}{a})
     \label{Gxx}\\
      \lim_{\bar{x}\rightarrow\bar{x}'}G_0(\bar{x},\,\bar{x}',\,m)=&\frac{i}{(2\pi)^2}\frac{1}{(a\rho)^2}\,\int_0^\infty \,d\omega\,\omega.
      \label{G0xx}
\end{align}
By substituting~\eqref{Gxx} in~\eqref{restcond}, we obtain
\begin{align}
     \pderivative{V^{\mathcal{R}}}{\varphi_0^2}\bigg|_{\varphi_0=0}&=\frac{m^2}{2}+\frac{\lambda}{16\pi^2}\,\frac{1}{(a\rho)^2}\,\int_0^\infty \,d\omega\,\omega\,\coth(\frac{\pi\omega}{a}) \label{consistency2}\\
     &=\frac{m^2}{2}+\frac{\lambda}{16\pi^2}\,\frac{1}{(a\rho)^2}\,\int_0^\infty \,d\omega\,\omega\,\biggl(1+\frac{2}{e^{\frac{2\pi\omega}{a}}-1}\biggr).
     \label{consistency}
\end{align}
We now explicitly note that taking $\rho=1/a$ considers points on the Rindler observer's worldline only, and gives a result identical to that of Ref.~\cite{Castorina:2012yg}, which in the reference is obtained through the resummation of Matsubara frequencies. The integrals in the above expressions diverge~\footnote{In particular, in~\eqref{consistency}, the divergence is confined to the first term of the integrand, which diverges quadratically.}, therefore regularization and renormalization procedures are necessary. We discuss them in the following section.

\section{Renormalization schemes\label{sec4}}

In order to renormalize the effective potential, i.e., to remove the divergences in~\eqref{consistency}, 
we define $m^2=m^2_R+\delta m^2$, where  $m^2_R$ is the renormalized squared-mass, and $\delta m^2$ is the mass counterterm. 
The renormalization procedure can be carried out in different ways, but only a frame-independent scheme yields a definition of $m^2_R$ which all observers agree on (let us say a \textit{universal} definition). If, on the contrary, frame-dependent counterterms are adopted, renormalized quantities are not universal anymore: this decisive fact has been overlooked in some past discussions of the physical outcomes of the theory under exam, and this is the fundamental reason why different conclusions exist about it in the literature.

In this section, we will discuss two possible renormalization schemes: scheme I, as considered, e.g., in~\cite{Castorina:2012yg}, where the counterterm is related to the Rindler vacuum state, and a second scheme, which we will call scheme II, in which the counterterm is obtained as a Minkowski-vacuum expectation value. In particular, in our discussion, we wish to point out, also conceptually, that the first option is frame-dependent, while the second is not. We also wish to clarify the related physical consequences. An explicit form for the counterterms will be computed in the next section, by making use of the geodesic point-splitting regularization method.
 
Let us start with the first choice I, which, as we already anticipated, is consistent, for example, with~\cite{Castorina:2012yg}. We have
\begin{equation}
    \label{count_term_1}
    \delta m^2_I=i\frac{\lambda}{2}\lim_{\bar{x}\rightarrow\bar{x}'}G_0(\bar{x},\,\bar{x}',\,m)=-\frac{\lambda}{2}\lim_{\bar{x}\rightarrow\bar{x}'}\bra{0_R}(\phi(\bar{x})\,\phi(\bar{x}'))_{\tau}\ket{0_R}.
\end{equation}
This choice, as apparent from~\eqref{G0xx}, cancels the quadratic divergence in~\eqref{consistency}, and yields
\begin{equation}
\begin{aligned}
    \pderivative{V^{\mathcal{R}}}{\varphi_0^2}\bigg|_{\varphi_0=0}&=\frac{{m^2_R}_I(\rho)}{2}+\frac{\lambda}{4}\lim_{\bar{x}\rightarrow\bar{x}'}(\bra{0_M}(\phi(\bar{x})\,\phi(\bar{x}'))_{\tau}\ket{0_M}-\bra{0_R}(\phi(\bar{x})\,\phi(\bar{x}'))_{\tau}\ket{0_R})\\
    &=\frac{{m^2_R}_I(\rho)}{2}+\frac{\lambda}{48}\,\biggl(\frac{1}{2\pi\rho}\biggr)^2.
\end{aligned}
     \label{criticaltempeq}
\end{equation}
The definition of the renormalized quantities, i.e. the renormalized mass in the above expression, involves the subtraction of a vacuum expectation value, which in the above expression is taken to be the Rindler vacuum state. Since points corresponding to a worldline with constant $\rho$ are associated with a Rindler observer undergoing a proper acceleration $a'=\rho^{-1}$, the second term in~\eqref{criticaltempeq} can be interpreted as a thermal correction (analogous to the case of a finite-temperature theory in inertial coordinates) at the Unruh temperature $T_{a'}=a'/2\pi$. Thus, the condition of symmetry restoration for $a$ larger than a critical acceleration, $a_c$, occurs only when~(\ref{criticaltempeq}) becomes positive. We emphasize that, if one insisted on considering the renormalized mass ${{m^2_R}_I}$ as constant (i.e., the same for each Rindler observer), then one would be led to conclude that such a critical acceleration would exist~\cite{Castorina:2012yg}, and, with it, symmetry restoration. But, let us go back to the renormalization scheme: we observe that the Rindler vacuum state is a concept that explicitly depends on the worldline $\rho=const$ we are considering (i.e. on the acceleration $a'$), which means that each Rindler observer has his/her own Rindler vacuum state. As a consequence, we can say that the renormalization scheme we have adopted here is \textit{frame-dependent}, because it comes with a mass counterterm depending on the observer, through the choice of $\rho$. So, the resulting renormalized squared-mass ${m^2_R}_{\mathrm{I}}$ hides a dependence on $\rho$. In other words, we can say that each Rindler observer, locally, measures a different ${m^2_R}_{\mathrm{I}}$. With this remark in mind, we conclude that, in \eqref{criticaltempeq}, when acceleration increases, also the ${{m^2_R}_I(\rho)}$ changes according to the renormalization scheme, and the \textit{symmetry restoration is not guaranteed} (in fact, as we will see, it does not take place).

A less intricate conclusion may be reached following a physical argument coming with a different choice of the renormalization scheme. In particular, we consider the mass counterterm defined as a Minkowski-vacuum expectation value (this is possibility II anticipated above):
\begin{equation}
    \label{count_term_2}
    \delta m^2_{\mathrm{II}}=i\frac{\lambda}{2}\lim_{\bar{x}\rightarrow\bar{x}'}G(\bar{x},\,\bar{x}',\,m)=-\frac{\lambda}{2}\lim_{\bar{x}\rightarrow\bar{x}'}\bra{0_M}(\phi(\bar{x})\,\phi(\bar{x}'))_{\tau}\ket{0_M}.
\end{equation}
This choice, as we can see from~\eqref{Gxx}, exactly cancels the integral in~\eqref{consistency2}, leading to
\begin{equation}
     \pderivative{V^{\mathcal{R}}}{\varphi_0^2}\bigg|_{\varphi_0=0}=\frac{{m^2_R}_{\mathrm{II}}}{2}.
     \label{criticaltempeq2}
\end{equation}
The definition of $\delta m^2$ made just above corresponds to the one used in the renormalization procedure performed in inertial frames. 
Since the Minkowski vacuum state is the same over the whole spacetime (in particular, on each worldline $\rho=const$), this choice is consistent with a \emph{frame-independent renormalization scheme}. This fact allows one to interpret ${m^2_R}_{\mathrm{II}}$ as a constant independent of the specific Rindler observer, including the limiting case $\rho\rightarrow\infty$ of an inertial observer. The physical outcome of this is that as long as ${m^2_R}_{\mathrm{II}}$ is a negative constant, there is no way for a Rindler observer to experience a symmetry restoration, no matter how big $a$ is, in agreement with~\cite{Unruh:1983ac}.

From the above discussion, we see that the details of the renormalization procedure, in any case, do not affect the answer to the question of whether a Rindler observer detects symmetry restoration for a super-critical acceleration, provided that one interprets the renormalized quantities correctly. The difference between the two approaches in the definition of the counter-terms $\delta m^2$ can be traced back to the choice of the vacuum state between the Rindler vacuum state (definition I), and the Minkowski vacuum state (definition II). With the former choice (I), corresponding to a frame-dependent renormalization prescription, the derivative of the effective potential \eqref{criticaltempeq} is the sum of the renormalized squared-mass, which is negative, and a finite (Unruh) temperature term, which is positive.
At this point one could be led to conclude that, by taking a sufficiently high acceleration, the second term would dominate on the first, thus realizing the symmetry restoration condition. But such an analysis would implicitly assume that the squared-mass term is the same for every accelerated observer, which is not the case. Indeed, a frame-dependent renormalization condition induces \textit{frame-dependent} renormalized quantities: in our case, the squared mass will depend on acceleration. In this way, one cannot vary acceleration without also affecting the squared-mass.

While from approach I it is not yet clear, this hidden acceleration-dependence, in fact, exactly cancels the Unruh term, and the truly frame-independent quantity is the derivative of the effective potential as a whole, which remains negative regardless of acceleration. This fact is immediately apparent following approach II. Being this second approach based on a frame-independent renormalization scheme, the resulting renormalized mass is constant (as measured by any Rindler observer). In this case, the finite temperature corrections cancel out in the one-loop effective potential \eqref{criticaltempeq2}, leaving the renormalized squared mass as the only contribution left: then we can conclude that symmetry restoration never takes place.

On this point, additionally, we have to comment that the effective action approach, used to determine the occurrence of a phase transition/symmetry restoration, has been defined in inertial systems, where renormalized quantities result to be the same for all the inertial observers. There is no reason, \emph{a priori}, to insist that such a procedure can be applied in the same way to a (uniformly) accelerated system.  Our result shows at a technical level how it is always possible to obtain a physically consistent interpretation in the two different frames, i.e., the inertial one and the uniformly accelerated one. This confirms that one can always support the conclusion in~\cite{Unruh:1983ac}, and ties different results obtained by other authors to the fact that a renormalization procedure based on the observer's vacuum state is \emph{formally} applied to an accelerated frame in complete analogy with inertial frames, without taking into account the dependence of the renormalized quantities on the acceleration\footnote{This situation, while possibly clouded by technical aspects, is not new in physics. In an inertial reference system Newton's equation equals the change in time of the linear momentum to the net \emph{active} force acting on the system. This procedure, however, is not the correct way to proceed in a non-inertial system, as the net active force measured in such a system \emph{needs} to be supplemented by inertial forces. In this sense, the descriptions of the \emph{same} system in the two \emph{different} frames are \emph{not conceptually equivalent}. The same reasoning could be applied to the case that we are considering here.}. Additionally, we wish to comment that, while in this case it is always possible to uphold the result obtained in an inertial system, we have this possibility only because spacetime respects a large symmetry group. In a more general situation, in particular, on a curved background, we would not have this way out of the problem. Also with this generalization in mind, it is important to understand in more detail the differences between schemes I and II. For instance, in the framework of a $\lambda\phi^4$ theory, such differences have been shown to play a role also in the context of secular growth \cite{Burgess:2018sou}. What we are going to do in the next section is to compute the two squared-mass counterterms explicitly, using a regularization method often applied to QFT in curved backgrounds: the covariant point-splitting method.

\section{Regularization and covariant point-splitting method\label{sec5}}

It is worthwhile to further explore the different nature and properties of the previously defined mass counterterms, which correspond to the two different and inequivalent renormalization schemes, I and II. At first, we are going to study these terms expressed as functions of propagators, and we will regularize them in the limit of vanishing geodesic distance (in what follows, the square of the geodesic distance will be denoted as $-2\sigma$). This regularization method goes under the name of \textit{covariant point-splitting} \cite{DeWitt:1975ys,Bunch:1977sq,Vilenkin:1982wt}. In this way, we will show that $\delta m^2_{\mathrm{I}}$ explicitly depends on the acceleration, while $\delta m^2_{\mathrm{II}}$ does not. Afterward, the form of such counterterms will be compared with the more general case of curved spacetime. It is worth noting that the point-splitting method is not the one and only covariant scheme, but it proves to be convenient when propagators are expressed in position space, as a function of geodesic distance.

First of all, we consider the well-known expression of the massive, scalar, quantum field propagator in Minkowski spacetime
 \begin{equation}
	\label{finalpropmink_pos}
	G\left(x,x'\right)=i\frac{m}{4\pi^2 \left(2\sigma\right)^{\frac{1}{2}}}K_1\left(m\left(2\sigma\right)^{\frac{1}{2}}\right)
\end{equation}
with $\left(2\sigma\right)^{\frac{1}{2}}=\sqrt{\left(\vb{x}-\vb{x^\prime}\right)^2-\left(t-t^\prime\right)^2}$, being the geodesic distance a covariant quantity, the propagator takes the same form in every coordinate system. In Rindler coordinates~\eqref{rindcoordtrasf} we can state that
 \begin{equation}
	G\left(\bar{x},\bar{x}'\right)=G\left(x,x'\right)
\end{equation}
and
$$
\left(2\sigma\right)^{\frac{1}{2}}=\sqrt{\rho^2+\rho^{\prime 2}-2\rho\rho^\prime \cosh\left(a\left(\tau-\tau^\prime\right)\right)+\left(\vb{x}_\perp-\vb{x}'_\perp\right)^2}.
$$
At this point, it is possible to obtain the regularized expression~(\ref{count_term_2}) for $\delta m^2_{\mathrm{II}}$ through series expansion of $G\left(\bar{x},\bar{x}'\right)$ for $\sigma \to 0$ as
 \begin{equation}
	G\left(\bar{x},\bar{x}'\right)=i\left(\frac{1}{8\pi^2\sigma}+\frac{m^2}{8\pi^2}\left[\gamma_E+\frac{1}{2}\log\left(\frac{m^2\sigma}{2}\right)\right]-\frac{m^2}{16\pi^2}\right)+ O\left(\sigma\right),
 \label{smallsigmaexpansion}
\end{equation}
where $\gamma_E$ is the Euler-Mascheroni constant.
To achieve the regularized expression~\eqref{count_term_1} for $\delta m^2_{\mathrm{II}}$, we need to start from the propagator evaluated with respect to the Rindler vacuum, in position space, i.e., $G_0\left(\bar{x},\bar{x}'\right)$ (\ref{finalproprind}). Refs.~\cite{Linet:1995mq} and~\cite{Candelas:1976jv} calculate it explicitly, by performing the integration over the momenta, and obtain
 \begin{equation}
\label{lin_can}
	G_0\left(\bar{x},\bar{x}'\right)=G\left(\bar{x},\bar{x}'\right)+i\frac{m}{8\pi^3}\int^{\infty}_{-\infty}du \,F\left(u,ia\tau-ia\tau'\right)\frac{K_1\left(m\,R_4(u)\right)}{R_4(u)},
\end{equation}
with
$$R_4(u)=\sqrt{2\rho\rho^\prime \cosh\left( u\right)+\rho^2+\rho^{\prime 2}+\left(\vb{x}_\perp-\vb{x}'_\perp\right)^2},$$
and
\begin{equation}
	F\left(u,\psi\right)=-\frac{\pi+\psi}{\left(\pi+\psi\right)^2+u^2}+\frac{\psi-\pi}{\left(\psi-\pi\right)^2+u^2}.
\end{equation}
Evaluating~\eqref{lin_can} in the coincidence limit, $\bar{x}\to\bar{x}'$, we obtain
 \begin{equation}
	\lim_{\bar{x}\rightarrow\bar{x}'} G_0\left(\bar{x},\bar{x}'\right)=\lim_{\bar{x}\rightarrow\bar{x}'} G\left(\bar{x},\bar{x}'\right)- i\frac{m}{4\pi^2}\int^{\infty}_{-\infty}\frac{du}{\pi^2+u^2} \frac{K_1\left(2m\rho'\cosh\left(\frac{u}{2}\right)\right)}{\sqrt{2\rho^{\prime 2}\left(1+\cosh(u)\right)}},
\end{equation}
where we have used the relation $\sqrt{1+\cosh\left(u\right)}=\sqrt{2}\cosh\left(u/2\right)$.

If we take the limit $m\rho\ll1$, as done before in order to obtain~\eqref{G0xx} and~\eqref{Gxx}, the integral can be calculated explicitly. This requirement means that the energy scale of acceleration, and the relative Unruh temperature, is much higher than the mass scale; this is exactly the typical behaviour that we assume for symmetries to be restored at, finite, high enough temperatures. Refs.~\cite{Castorina:2012yg},~\cite{Ebert:2006bh} and~\cite{Ohsaku:2004rv} assume the same limit.
In this way, we can perform the following expansion for small $m\rho$
  \begin{equation}
 K_1\left(2m\rho'\cosh\left(\frac{u}{2}\right)\right)=\frac{1}{2m\rho'\cosh\left(\frac{u}{2}\right)}+O\left(m\rho'\right),
\end{equation}
and the propagator in the coincidence limit becomes
 \begin{equation}
	\lim_{\bar{x}\rightarrow\bar{x}'} G_0\left(\bar{x},\bar{x}'\right)=\lim_{\bar{x}\rightarrow\bar{x}'} G\left(\bar{x},\bar{x}'\right)-\frac{i}{8\pi^2\rho^2}\int^{\infty}_{-\infty}\frac{du}{\pi^2+u^2}\frac{1}{1+\cosh\left(u\right)}.
\end{equation}
It is possible to further simplify this expression by evaluating the integral on the RHS,
 \begin{equation}
\int^{\infty}_{-\infty}\frac{du}{\pi^2+u^2}\frac{1}{1+\cosh\left(u\right)}=\frac{1}{6},
\end{equation}
to finally obtain
 \begin{equation}
	\lim_{\bar{x}\rightarrow\bar{x}'} G_0\left(\bar{x},\bar{x}'\right)=\lim_{\bar{x}\rightarrow\bar{x}'} G\left(\bar{x},\bar{x}'\right)-\frac{i}{48\pi^2\rho^2}.
\end{equation}

Summarizing the results obtained so far and using the expansion \eqref{smallsigmaexpansion}, we can write the mass counterterms in~\eqref{count_term_1} and~\eqref{count_term_2} as
\begin{align} \label{count_f_1}
  \delta m^2_{\mathrm{I}}&=-\frac{\lambda}{2}\left(\frac{1}{8\pi^2\sigma}+\frac{m^2}{8\pi^2}\left[\gamma_E+\frac{1}{2}\log\left(\frac{m^2\sigma}{2}\right)\right]-\frac{m^2}{16\pi^2}-\frac{1}{48\pi^2\rho^2}\right),\\
 \label{count_f_2}
  \delta m^2_{\mathrm{II}}&=-\frac{\lambda}{2}\left(\frac{1}{8\pi^2\sigma}+\frac{m^2}{8\pi^2}\left[\gamma_E+\frac{1}{2}\log\left(\frac{m^2\sigma}{2}\right)\right]-\frac{m^2}{16\pi^2}\right).
\end{align}
We observe that, whereas choice II is frame-independent, in choice I, observer-dependence arises because of the last term: as we discussed above, the counterterm changes according to the wordline $\rho=1/a$ that one considers.  

At the very least, we want to compare these expressions with a set of counterterms derived with the geodesic point-splitting method in the more general case of curved spacetime in Ref.~\cite{Christensen:1976vb}. Such a method is one of the most popular and used regularization procedures in curved spacetimes, because it allows to express divergent terms in a covariant form.
In curved spacetime, the divergent limit $\lim_{x\rightarrow x'}G_{\mathrm{CS}}(x,\,x')$ of the propagator is regularized as
\begin{align}
\lim_{x\rightarrow x'}G_{\mathrm{CS}}\left(x,x'\right)=&i\biggl(\frac{1}{8\pi^2\sigma}+\frac{m^2+\left(\xi-\frac{1}{6}\right)R}{8\pi^2}\left[\gamma_E+\frac{1}{2}\log\left(\frac{m^2\sigma}{2}\right)\right]+ \notag\\
&-\frac{m^2}{16\pi^2}+\frac{1}{96\pi^2}R_{\alpha\beta}\frac{\sigma^\alpha\sigma^\beta}{\sigma}\biggr),
\label{Christensen}
\end{align}
where $R_{\alpha\beta}$ is the Ricci tensor, $R$ is the Ricci scalar, $\xi$ is the coupling constant to gravity, and ${\sigma^\alpha}$ are the components of the vector field tangent to the geodesic, oriented in the $x \to x'$ direction and have a length equal to the geodesic distance. In flat spacetime, which is our case, Ricci-tensor terms in~\eqref{Christensen} vanish, and using it to calculate the mass counterterms yields
\begin{equation}
\delta m^2_{\mathrm{CS}}=-\frac{\lambda}{2}\left(\frac{1}{8\pi^2\sigma}+\frac{m^2}{8\pi^2}\left[\gamma_E+\frac{1}{2}\log\left(\frac{m^2\sigma}{2}\right)\right]-\frac{m^2}{16\pi^2}\right),
\end{equation}
which is exactly~\eqref{count_f_2}. 
We conclude that choice II is the correct one to preserve covariance, while choice I introduces a frame-dependent term that spoils this property.

\section{Conclusion\label{sec6}}

We have revisited the analysis of a scalar field theory with spontaneous symmetry breaking in a Minkowski spacetime from the point of view of an accelerated observer. The issue of accelerated observers experiencing or not a symmetry restoration above a critical proper acceleration has been dealt with in a seminal work by Unruh and Weiss, who concluded with a negative answer, motivated by the invariance properties of vacuum Green functions. We have recovered the same result using the effective action method, where the issue of renormalization is crucial. We have shown that the counterterms that one defines are vacuum expectation values and that the choice of the vacuum state between the Rindler vacuum and the Minkowski vacuum leads to different renormalization schemes, reflecting the different invariance properties of the two vacua. While the Rindler vacuum brings to a frame-dependent scheme, whose physical interpretation can be misleading, the Minkowski vacuum corresponds to a frame-independent renormalization scheme, whose natural consequence is to rule out any symmetry restoration, in agreement with the conclusion by Unruh and Weiss.

We note that our approach considers a given system, and the ``accelerated'' case corresponds to an accelerated observer that performs measurements on this system. We do not consider more sophisticated setups, e.g. those of systems with moving boundaries, or those where the relative acceleration is obtained by switching on some interaction between the system and the observer. In particular, exponentially receding mirrors as viewed by stationary observers, have been shown to share similarities with radiating black holes, in an even stronger physical sense than the Unruh effect~\cite{Davies:1977yv}. It is possible that the conclusion in this work may have to be reconsidered in these more general setups, i.e., in these cases the acceleration could potentially induce symmetry restoration, at least in principle. The technical subtleties in treating the problem are exemplified in this work by the two possible renormalization schemes that we have considered. Of course, it would be interesting to extend this analysis to the case in which general covariance is considered as the relevant symmetry from the beginning. Indeed, while it is natural that, if we start with global Lorentz invariance, there has always to be a way to reconcile the, possibly different, physical interpretations of the accelerated observers with that of the inertial ones, the situation changes drastically, e.g., in a genuinely curved spacetime. In this case, in general, it may not be possible to define a privileged class of reference frames (and this is, in fact, the reason why general covariance is the appropriate setup in this more general case) and, indeed, the effect of curvature on symmetry breaking/restoration lifts the problem to a completely different level. We think, however, that our analysis shows a reasonable approach that could be considered to develop a consistent treatment of these more complex situations, as, for instance, theories involving gauge fields or spinor fields, also on curved backgrounds.

\section*{Acknowledgements}

One of us, SA, would like to thank for hospitality the Theoretical Astrophysics Group at the Department of Physics of Kyoto University, where several ideas present in this work were firstly conceived. AF acknowledges the support of the Japanese Society for the Promotion of Science Grant-in-Aid for Scientific Research (KAKENHI, Grant No. 21K03540).

\providecommand{\href}[2]{#2}\begingroup\raggedright\endgroup

\end{document}